\documentstyle[11pt]{article}

        \newcounter{sect}

        \newcommand{\be}{\begin{equation}}
        \newcommand{\ee}{\end{equation}}
        \newcommand{\bea}{\begin{eqnarray}}
        \newcommand{\eea}{\end{eqnarray}}
\newcommand{\ben}{\begin{displaymath}}
        \newcommand{\een}{\end{displaymath}}
        \newcommand{\bean}{\begin{eqnarray*}}
        \newcommand{\eean}{\end{eqnarray*}}
        \newcommand{\nno}{\nonumber \\}

        \newtheorem{theorem}{Theorem}
        \newtheorem{definition}{Definition}
\newtheorem{lemma}{Lemma}
\newtheorem{corollary}{Corollary}
\newtheorem{proposition}{Proposition}
 \newcommand{\A}{{\cal A}}
        \newcommand{\D}{{\cal D}}
        \newcommand{\HH}{{\cal H}}

        \newcommand{\B}{{\cal B}}

        \newcommand{\s}{\sigma}

        \newcommand{\dd}{|\D|}
        \newcommand{\n}{\parallel}
        \newcommand{\R}{\mathbf R}

\newcommand{\Z}{\mathbf Z}
\newcommand{\proof}[1]{\noindent \normalfont{\textbf{Proof} \  \  #1 \hfill $\Box$} \par}

\newcommand{\bma}{\left(\begin{array}{cc}}
\newcommand{\ema}{\end{array}\right)}

\newcommand{\tto}{\longrightarrow}

\newcommand{\text}[1]{\mbox{\normalfont #1}}

\newcounter{propcount}
\begin{document}
\title{Poincar\'{e} Duality and Spin$^c$ Structures for Complete Noncommutative
Manifolds}
\author{A. Rennie}
\maketitle
\centerline{Abstract}

We prove a noncompact Serre-Swan theorem characterising modules
which are sections of vector bundles not necessarily trivial at
infinity. We then identify the endomorphism algebras of the
resulting modules. 
The endomorphism results continue to hold for the generalisation of
these modules to noncommutative,
nonunital algebras. Finally, we apply these results to not necessarily compact 
noncommutative spin manifolds, proving that Poincar\'{e} Duality 
implies the Morita equivalence of the `algebra of functions' and the 
`Clifford algebra.'

{\bf AMS 2000 Subject Classification} 46L08 and 46L89; also 46L80

\section{Introduction}

Part of the folklore of noncommutative geometry is that the Serre-Swan theorem
generalises to noncompact spaces. It is also generally assumed (and sometimes
stated) that this generalisation
introduces essentially nothing new. In section 2, we show that there is a
nontrivial 
characterisation of
the sections of vector bundles tending to zero at infinity on
locally 
compact spaces.

Let $C_0(X)$ be the continuous functions on $X$ which vanish at
infinity. Then the main result of the next section is that a
$C_0(X)$ module $E$ is isomorphic to the bundle of sections vanishing
at infinity on some vector bundle $V\to X$ if and only if 
\be E\cong pC_0(X)^n,\qquad p^*=p^2=p\in M_n(C(X^c)),\ee
where $X^c$ is some compactification of $X$. Dealing with general compactifications
allows one to speak about bundles not trivial at infinity.

If $A$ is a nonunital $\s$-unital $C^*$-algebra, then we say that
an $A$-module $E$ is $A_b$ finite projective if $E=pA^n$ for some
projection $p\in M_n(A_b)$ where $A_b$ is some unitization of
$A$ (we write $A_b$ in analogy with a subalgebra of bounded
functions). 
We identify such modules as the noncommutative analogue of 
sections of vector bundles vanishing at infinity. With this level
of generality we identify the compact endomorphism and the full
endomorphism algebras of these modules. We find that
\be End_A(E)=pM_n(A_b)p\qquad End_A^0(E)=pM_n(A)p.\ee
In particular, the compact endomorphism algebra of such a module
is nonunital.

In the last section we apply these results to noncompact versions
of Connes' noncommutative spin manifolds. This requires
generalisations of his axioms which are not presented here in
full; see \cite{AN}. We describe the results and assumptions
necessary for us to show that, with $\HH_\infty$ the right pre-$C^*$
$\A$-module associated to the manifold,  
\be \Omega^*_\D(\A)\cong End^0_\A(\HH_\infty)\qquad
\Omega^*_\D(\A_b)\cong End_\A(\HH_\infty).\ee
Here $\A_b$ is a unitization of $\A$ and the algebra
$\Omega^*_\D(\A)$ is obtained by representing the universal
differential algebra of $\A$. Similar comments apply to
$\Omega^*_\D(\A_b)$,
but we can also obtain it as the `smooth strong' closure of
$\Omega^*_\D(\A)$, \cite{AN}. These results show that $\A$ and 
$\Omega^*_\D(\A)$ are strongly Morita equivalent. From the description of spin$^c$
structures on a manifold as 
Morita equivalence bimodules between the algebra of functions and the complex Clifford
algebra, \cite{RP}, and the fact that $\Omega^*_\D(\A)$ coincides with the Clifford
algebra in the commutative case, we see that these isomorphisms provide a
noncommutative definition of spin$^c$ structure. This point of view is strengthened by
 the module $\HH_\infty$ coinciding with the smooth sections of the corresponding complex 
 spinor
bundle in the commutative case, \cite{C1,A,AN}.

\section{The Nonunital Serre-Swan Theorem}

The first result concerns modules finitely generated over the
nonunital 
algebra $A$, and
projective over $A^+$. These have occasionally been touted as the
analogues of sections of vector bundles vanishing at infinity. 
Essentially we characterise them in order
to 
dispense with them.

\begin{lemma} Let $A$ be nonunital. Then $E$ is a finitely
  generated (right) 
$A$-module
which is projective over $A^+$ if and only if $E$ is of the form
$pA^n$, 
where $p=p^2\in
M_n(A)$.

\medskip
\proof{Since $E$ is a finitely generated and projective $A^+$-module,
\be E\cong p(A^+)^n,\ee
where $p\in M_n(A^+)$. However, $E$ is finitely generated over $A$, so for some
$M$ there exists $e_1,...,e_M\in E$ such that for all $e\in E$ there are $a_1,...,a_M\in A$
with 
\be e= e_1 a_1+\cdots e_M a_M.\ee
So in fact, as $A$ is an ideal in  $A^+$, every $e\in E$ is an element of $pA^n$. In particular, the
elements
\be p\left(\begin{array}{ccccc}1\\ 0\\ \cdot\\\cdot\\ 0\end{array}\right)=
\left(\begin{array}{ccccc}p_{11}\\ p_{21}\\ \cdot\\\cdot\\ p_{n1}\end{array}\right),
\ \cdots\ ,
p\left(\begin{array}{ccccc}0\\ 0\\ \cdot\\\cdot\\ 1\end{array}\right)=
\left(\begin{array}{ccccc}p_{1n}\\ p_{2n}\\ \cdot\\\cdot\\ p_{nn}\end{array}\right),\ee
are contained in $E$, and so $p\in M_n(A)$.
\smallskip\newline
Conversely, if $E=pA^n=p(A^+)^n$, $p\in M_n(A)$, then $E$ is projective over $A^+$ since
\be E\oplus (1-p)(A^+)^n=(A^+)^n\ee
is a free $A^+$-module. It is finitely generated over $A$, because the column vectors
$p_j=(p_{ij})$, $j=1,...,n$, above provide a system of generators.}
\end{lemma}

\noindent To continue the investigation, we note that if $i:A\hookrightarrow A_b$ 
is a unitization of
$A$, we can define the pull-back of a right $A_b$-module $E$ by
\be i^*E=E|_A:=Ei(A)=\{ei(a): e\in E,\ a\in A\},\ee
with the obvious right action of $A$ on $E|_A$. Note that as an $A_b$-module, $E|_A$ is a 
submodule of $E$. In fact, the pullback 
works for 
any embedding
$i:A\hookrightarrow A_b$ of $A$ as an ideal in $A_b$, even if it is not closed.
\smallskip\newline
If
$i: X\hookrightarrow X^c$ is an embedding of the locally compact 
Hausdorff space $X$ in
the compact Hausdorff space $X^c$ (automatically as a dense open
subset), 
then we may
define a unitization of $C_0(X)$ as follows. Define, \cite{RW},
\be i_*:C_0(X)\hookrightarrow C(X^c)\ee
by
\be (i_*f)(y)=\left\{\begin{array}{ll} 0 & 
\mbox{ if }y\not\in i(X)\\ f(x) & \mbox{ if }y=i(x),\
x\in X.\end{array}\right.\ee
\medskip\newline
{\bf Example} Let $1\to X^c$ be the trivial line bundle over 
$X^c$, where $X^c$ is a
compactification of the locally compact noncompact
Hausdorff space $X$. The space of sections of this line bundle is 
$E=C(X^c)$, and this is 
a right
$C(X^c)$ module. We have the unitization map, 
$i:C_0(X)\hookrightarrow C(X^c)$, and so
we can pull $E$ back to $C_0(X)$. We find
\be i^*E=i^*C(X^c)=C(X^c)i(C_0(X))=C_0(X).\ee
Thus we obtain the space of sections vanishing at infinity. 
\medskip\newline
{\bf Example} Let $X$ be as above, and consider the embedding 
$i:C_c(X)\hookrightarrow
C_0(X)$, where $C_c(X)$ is the algebra of compactly supported
functions. 
If $E=\Gamma_0(X,V)$ is the $C_0(X)$-module of sections vanishing at infinity of
the vector bundle $V$, then we can pull it back to $C_c(X)$ as above. Then
\be i^*E=\Gamma_c(X,V)\ee
and we obtain the compactly supported sections. In this case the resulting module is only
a pre-$C^*$-module, because $C_c(X)$ is not a $C^*$-algebra.

\begin{definition} Let $A$ be a $\s$-unital $C^*$-algebra. If 
$i:A\hookrightarrow A_b$ is a unitization, we say that an $A$-module
is $A_b$ finite projective if it is of the form $i^*E$ for some 
finite projective
$A_b$-module $E$.
\end{definition}
\noindent
With this definition we are ready to state the main result of
this section.

\begin{theorem}[Nonunital Serre-Swan]
Let $X$ be a locally compact Hausdorff space,
$A=C_0(X)$, and $A_b=C(X^c)$ for some compactification $X^c$ of
$X$. 
Then a right
$A$-module $E$ is of the form $E=pA^n$, $p\in M_n(A_b)$ a
projection, 
if and only if
$E=\Gamma_0(X,V|_X)$, where $V\to X^c$ is a vector bundle and 
$\Gamma_0$ denotes
 the sections vanishing at infinity.
  
\medskip
\proof{Suppose that $E=\Gamma_0(X,V|_X)$, where $V\to X^c$ is a 
vector bundle. Then
for some $n$ and projection $p\in M_n(A_b)$, 
\be \Gamma(X^c,V)\cong p(A_b)^n,\ee
by the Serre-Swan theorem, \cite{S}. Note that as $X\subset X^c$
is dense 
and open, and rank $p$
is locally constant, $p\not\in M_n(A)$. Setting
$i:A\hookrightarrow A_b$ to 
be the
unitization, we have
\be i^*\Gamma(X^c,V)=pA^n=\Gamma_0(X,V|_X).\ee
Conversely, let $E=pA^n$ with $p\in M_n(A_b)$ a projection. Then
we 
can define a finite
projective $A_b$-module, $\tilde E=p(A_b)^n$, with the obvious
right 
action of $A_b$. By
the Serre-Swan theorem, there is a vector bundle $V\to X^c$ such that 
\be \tilde E\cong p(A_b)^n=\Gamma(X^c,V).\ee
Employing the pull-back by the unitization map as above immediately shows that
\be i^*\tilde E=E=\Gamma_0(X,V|_X).\ee}
\end{theorem}

\begin{corollary} With $A$ and $A_b$ as above, the $A$-module $E$
  is 
isomorphic to
$\Gamma_0(X,V)$, for $V\to X$ trivial at infinity, if and only if 
$E\cong pA^n$ for some 
$p\in M_n(A^+)$.
\end{corollary}
\noindent
In fact, since a bundle trivial at infinity will extend to any
compactification (trivially), there must be $p\in M_n(A_b)$ such that
$\Gamma_0(X,V)=pA^n$, for any unitization $A_b$. This does not contradict the corollary.
Excision in $K$-theory, \cite{RH}, tells us that the compactly supported $K$-theory
(the usual definition in the nonunital/noncompact case) is independent of any
compactification. However, the nonunital Serre-Swan theorem is telling us that to get
a handle on the actual vector bundles, and not just the resulting cohomology theory,
we need to take account of all compactifications/unitizations simultaneously.
Fortunately, the existence of a 
maximal compactification ensures the existence of a
maximal compactification to which a given bundle extends. Note that 
any notion of bounded
cohomology (the natural dual of singular homology defined using finite chains) in
$K$-theory will require this notion. 
\medskip\newline
Also, it is now clear why we are not interested in
modules of the form $pA^n$, where $A$ is nonunital and $p\in M_n(A)$. These do not
correspond to sections vanishing at infinity in the commutative case; they do not even
characterise a vector bundle on $X^+$, since we do not have locally constant rank at
infinity. It would be interesting to see what, if any, geometric content these 
modules have
in the commutative case. 

{\bf Example} If $X$ is the interior of a manifold with boundary $\overline{X}$, then
the sections of the tangent bundle of $X$ vanishing on the boundary is certainly an
example of the above phenomena. This seems somewhat trivial as we have a compact space
to which the vector bundle extends, and so we can realise it in the usual compact
Serre-Swan fashion. 
\medskip\newline
A more genuine seeming example is any (finite dimensional) manifold which does not
have finitely generated cohomology groups. It is easy to construct a vector bundle $V$ 
on such a space which is not trivial outside any compact set. An example of such a
space would be ${\R}^n$ with the balls centred at integer coordinates 
of radius $\frac{1}{4}$ deleted. It is easy to construct bundles on this space which
are not trivial outside any compact set.

We now drop the assumption that $A$ is commutative, and describe
the 
endomorphism
algebras of $A_b$ finite projective $A$-modules. By results 
in \cite{V, L}, finite projective
$A_b$-modules $E$ can be regarded as $C^*$-modules once an 
$A_b$-Hermitian form is 
chosen.
This form automatically restricts to the $A_b$-submodule $E|_A$, 
since $A$ is an ideal in
$A_b$. It is the
endomorphism algebras of these modules that we now describe. 
This is relevant for the 
noncommutative geometry of nonunital algebras, \cite{C}.

\begin{proposition} Let $E|_A$ be an $A_b$-finite projective
  $A$-module. 
Then, as a
$C^*$ $A$-module, we have
\be End^0_A(E|_A)=pM_n(A)p\qquad End_A(E|_A)=pM_n(A_b)p,\ee
where $End^0_A(E|_A)$ denotes the compact endomorphisms of $E|_A$ and 
$End_A(E|_A)$ denotes the $C^*$-algebra of adjointable operators on $E|_A$. 

\medskip
\proof{The $A_b$-module $E$ is of the form $E=p(A_b)^n$. Writing
  $M(B)$ 
for the
multiplier algebra of a $C^*$-algebra $B$, it is then standard that,
\cite{V,L},
\be End_{A_b}(E)\cong M(End^0_{A_b}(E))=End^0_{A_b}(E)\cong pM_n(A_b)p.\ee
These equalities follow from $End^0(E)$
being unital, which is itself a standard result.

To prove the above assertions, we begin with a variant on a standard 
isomorphism, \cite{V}. 
Denoting the
$A$-finite rank operators on $E|_A$ by $End^{00}_A(E)$, define
$\theta:End^{00}_A(E|_A)\tto pM_n(A)p$ by setting
$\theta(|p\xi)(p\eta|)$ 
to be the matrix
with $i,j$-th entry $\sum_{k,l}p_{ik}\xi_k\eta^*_lp_{lj}$. One
checks that 
this is a
$*$-homomorphism and an isometry and has dense range. As such,
$\theta$ 
extends to a
$*$-isomorphism of $End^0_A(E|_A)\cong pM_n(A)p$.
\smallskip\newline
As $i^*E$ is a right $A_b$-submodule of $E$, we see that
\be End_{A_b}(i^*E)=End_A(i^*E).\ee
Also, $i^*E$ is a left $End_{A_b}(E)$-submodule, as is easily
checked. 
Consequently
\be End_A(i^*E)=End_{A_b}(E)=pM_n(A_b)p.\ee}
\end{proposition}

Note that in \cite[Appendix A]{RH}, there is a different approach, and elements of
$End^0_A(E|_A)$ are called endomorphisms carried by $A$. Modules of the above type and
their endomorphism algebras are used to describe elements of relative $K$-theory. 

\section{Complete Noncommutative Manifolds}

In \cite{C1,A,V}, closed noncommutative spin manifolds were defined
as those spectral triples satisfying certain additional
axioms. Moreover, these were shown to reduce to ordinary closed
spin manifolds in the commutative case. 
\smallskip\newline
In \cite{AN}, this is
generalised to complete (i.e. not necessarily compact but
geodesically complete and without boundary) spin manifolds. The
extension of the noncompact axioms to the noncommutative case
makes sense, and the resulting objects are called complete
noncommutative spin manifolds. 
\smallskip\newline
Note that this extension is
nontrivial in several regards, there being algebraic, analytic
and homological problems. The chief algebraic problem is the
appropriate characterisation of the sections of vector bundles
on noncompact spaces. This was addressed in the previous section, 
and in this section we apply these results to complete
noncommutative manifolds. 

In addition, to prove that one recovers complete noncompact manifolds in the commutative
case, one requires that the nonunital algebra of functions is local, in that it possesses
a dense subalgebra which has local units. We will not need any particular details of
these algebras, other than knowing that the constructions and definitions we present make
sense for these algebras; see \cite{AN}. Also, a unital algebra is necessarily local. 

We prove two main results. The first is that the (pre) Hilbert
space appearing in the definitions provides a (pre) strong Morita
equivalence between $\A$ and $\Omega^*_\D(\A)$, the analogues of
the algebra of functions and Clifford algebra respectively. The
second result is a corollary of the first, and states that
$\Omega^*_\D(\A)$ is finite projective over $\A$ as a left or
right module. This shows that it behaves like the sections of a
bundle. We begin by introducing the basic objects of noncommutative geometry.

\begin{definition}
A spectral triple $(\A,\HH,\D)$ is given by 

1) A representation $\pi:\A\to\B(\HH)$ of a local algebra $\A$ on the Hilbert space
$\HH$.

2) A closed, (unbounded) self-adjoint operator $\D:\mbox{dom}\D\to\HH$ such that 
$[\D,a]$ extends to a bounded operator on $\HH$ for all $a\in\A$ and 
$a(1+\D^2)^{-\frac{1}{2}}$ is compact for all $a\in\A$.

The triple is said to be even if there is an operator $\Gamma=\Gamma^*$ such that 
$\Gamma^2=1$, $[\Gamma,a]=0$ for all $a\in\A$ and $\Gamma\D+\D\Gamma=0$ (i.e.
$\Gamma$ is a ${\Z}_2$-grading such that $\D$ is odd and $\A$ is even.) Otherwise the
triple is called odd.
\end{definition}
\noindent
Since $\A$ is represented on Hilbert space we may unambiguously speak about the norm
on $\A$, and the norm closure $\overline{\A}=A$.
\medskip\newline
We recall the basic setup, and refer to \cite{A} for the
details. Given a spectral triple $(\A,\HH,\D)$, we say that it is
a complete (noncommutative) spin manifold if it satisfies the axioms in
\cite{AN}. There are several of these axioms and their
consequences that we shall require. 
\medskip\newline
1) There is a unitization $\A_b$ of $\A$ such that the smooth domain of $\D$,
\be \HH_\infty=\bigcap_{m\geq 1}\ \mbox{dom} \D^m,\ee
is an $\A_b$ finite projective (right) $\A$-module, and $\A$ is a
Fr\'{e}chet algebra for the seminorms provided by
\be q_n(a)=\n\delta^n(a)\n,\quad \delta(a)=[\dd,a].\ee
\smallskip\newline
One can show that $\A_b$ is also complete for the strong topology
determined by these seminorms, and is in fact characterised as the strong closure of
$\A$ with respect to these seminorms. 
\medskip\newline
2) The first order condition holds. This means that $\HH_\infty$ is
in fact an $\A$-bimodule, or equivalently an $\A\otimes\A^{op}$ module, and
\be [a,b^{op}]=0\quad[[\D,a],b^{op}]=0,\ \forall a,b\in\A.\ee
\smallskip\newline
This ensures that the  representation of  the universal
differential algebra of $\A$, $\Omega^*(\A)$, given by
\be
\pi(a\delta(b_1)\cdots\delta(b_k))=a[\D,b_1]\cdots[\D,b_k],\ee
is contained in the commutant of $\A^{op}$, \cite{C1,A,AN}. As the above
conditions are symmetric in $\A$ and $\A^{op}$, a similar
`opified' statement holds also. Moreover, all the above comments
hold for $\A_b$. One can also show that $\Omega^*_\D(\A)$ is contained in 
the smooth domain of the derivation $\delta$.
\smallskip\newline
An alternative approach is to regard $\HH_\infty$ as a pre-$C^*$ $\A$-module. From
this point of view, $\Omega^*_\D(\A)$ is contained in $End_\A(\HH_\infty)$. This 
is the approach we will adopt here. 
\medskip\newline
3) Most importantly, the spectral triple satisfies Poincar\'{e}
Duality. The $K$-homology class defined by the above spectral triple
\ben \mu\in K(A\otimes A^{op}),\een 
provides us with a map 
\ben \bigcap\mu:K_*(A)\tto K^*_c(A).\een
We say that $(\A,\HH,\D)$ satisfies Poincar\'{e} Duality if this map 
is an isomorphism. Here $K^*_c(\A)$ is $K$-homology with compact supports. This group
can not be defined for all smooth, nonunital algebras, and imposes a tight restriction
on the algebras under consideration, \cite{AN}. 
\smallskip\newline
We have written the map as a cap product which
is not strictly true in the noncommutative case, but the Kasparov
product gives us a map more properly called a slant map,
and this is how we compute it; see \cite{C,C1,A,AN,RH}. 
For information on $K$-homology and the various products,
see \cite{RH}.
\smallskip\newline
Our aim for the rest of this section is to show that
$\Omega^*_\D(\A)$ and $\A$ are strongly Morita equivalent. 
To do this,
we recall that sections of the spinor bundle form an irreducible 
representation of the
Clifford algebra in the sense that it is irreducible 
(in the usual sense of the word
irreducible) fibrewise. The following concept is useful in
dealing with this 
situation.

\begin{definition} Let $E$ be a right $C^*$ $A$-module. Then we say that the
representation $\pi:B\to End_A(E)$ is $A$-irreducible if the only
operators 
in $End_A(E)$
commuting with $B$ are the scalars. 
\end{definition}
\noindent
{\bf Remark} This works just as well for pre-$C^*$-modules and 
pre-$C^*$-algebras.
\smallskip\newline
{\bf Example} Let $E=\Gamma(X,V)$ be the sections of a vector
bundle $V$ over the closed Riemannian spin manifold $X$, and suppose that $E$ is
a Clifford bundle. That is, $E$ admits a (fibrewise)
representation of the sections of the Clifford algebra of the
cotangent bundle of $X$. Then $E$ is a $C^\infty(X)$
pre-$C^*$-module, and the representation
\be Cliff(T^*X)\to End_{C^\infty(X)}(E)\ee
is irreducible if it is so fibrewise. For if $B:E\to E$ is
$C^\infty(X)$-linear, $B\in
End(\Gamma(X,V))=\Gamma(End(X,V))$. Thus $B$ is a smooth
matrix-valued function on $X$, and if it is to commute with
$Cliff(T^*X)$, it must do so fibrewise. Hence the representation 
of $Cliff(T^*X)$ is
$C^\infty(X)$ irreducible if and only if the representation 
of $Cliff(T^*_xX)$ is irreducible
on $E_x$ for all $x\in X$. 
\bigskip\newline
To compare this notion of irreducible with others we have the
following two 
results.

\begin{lemma} The representation $\pi:B\to End_A(E)$ is
  irreducible if and 
only if $B$
has no invariant complementable submodules in $E$.

\medskip
\proof{Suppose that $V\subseteq E$ is a complementable submodule with 
$BV\subseteq V$. Then, since complementable submodules are
precisely those 
which are
the images of projections in $End_A(E)$, there exists a
projection 
$Q\in End_A(E)$ such that $QE=V$ and
$Q$ commutes with $B$.
\medskip\newline
Conversely, suppose there exists $T\in End_A(E)$ such that
$Tb=bT$ for all 
$b\in B$.
Then $Image T$ is a complementable submodule, being the image of
an 
adjointable map,
and $B Image T\subseteq Image T$.}
\end{lemma}

\begin{corollary} The representation $\pi:B\to End_A(E)$ is
  irreducible if 
and only if there
do not exist projections $P,Q\in End_A(E)$ such that $E=PE\oplus QE$ and $P,Q$
commute with $B$.
\end{corollary} 
With these tools in hand we can prove our main technical result. The hypotheses
apparently 
make the result somewhat obvious and perhaps trivial, but the proof 
shows that there is much
more at work.

\begin{theorem} Suppose that $(\A,\HH,\D)$ is a noncommutative
spin manifold, and that the only bounded operators commuting with $\A$ and
$\D$ on $\HH$ are scalars. Then, regarding  $\HH_\infty$ as a pre-$C^*$-$\A$-module, the
representation of $\Omega^*_\D(\A_b)$ is $\A$-irreducible. 

\medskip
\proof{Suppose not. We begin by writing $\HH_\infty=p\A^n$ for some projection $p\in
M_n(\A_b)$. Then there exists $e=e^*=e^2\in pM_n(\A_b)p$ such that 

1) $[e,p]=0$, since $e\in pM_n(\A_b)p$, and 

2) $[\omega,e]=0$ for all $\omega\in\Omega^*_\D(\A_b)$, by hypothesis.
\medskip\newline
We write 
\bean \D & = & e\D e+(1-e)\D(1-e)+e\D(1-e)+(1-e)\D e\nno
& = & e\D e+(1-e)\D(1-e)+B\nno
& = & \D_e + B.\eean
Using 2) we have that for all $a\in\A_b$
\ben [\D,a]=e[\D,a]e+(1-e)[\D,a](1-e)=[\D_e,a]\een
so $[B,a]=0$. In other words, $B=B^*$ is $\A_b$-linear 
and so can easily be seen to be bounded (just check its behaviour on a generating set). 
Consequently,
the map
\ben t\tto\D_e+tB\een
provides us with an operator homotopy from $\D$ to $\D_e$. It is easy to check that
this homotopy preserves the $K$-homology class of $\mu$, \cite{RH}. Hence we may write
\bean \mu=[(\A,\HH,\D)] & = & [(\A,e\HH,e\D e)]+[(\A,(1-e)\HH,(1-e)\D(1-e))]\nno
& \in & K^*(\A\otimes\A^{op}).\eean
The next step is to show that $[e]=[Id_\HH-e]=[p-e]=[p]-[e]$ in 
the $K$-theory of $\A_b$. To do this we will construct an explicit Murray-von Neumann
equivalence. First we note that 
\ben [\D,e]=[B,e]=(1-e)\D e-e\D(1-e),\een
and using $e^2=e$, $(de)e=(1-e)(de)$ et cetera, we compute
\bean \D de & = & \D(1-e)\D e-\D e\D(1-e)\nno
& = & [\D,(1-e)](1-e)[\D,e]-[\D,e]e[\D,(1-e)]\nno
& = & -e[\D,e][\D,e]+(1-e)[\D,e][\D,e]\nno
& = & -(2e-1)dede.\eean
In a completely analogous fashion we find that
\ben de\D=(2e-1)dede.\een
This shows that $\D de=-de\D$ and so $\D dede=dede\D$. Hence $[\D,e][\D,e]$ commutes 
with all
$a\in\A$ and $\D$, and so must be a scalar. Moreover, $(de)^*=-(de)$ so
\ben -dede=B^2\geq 0\een
is a positive real number. Suppose first that $m=0$. Then $de=0$ so $e$ is a scalar,
and so the representation of $\Omega^*_\D(\A_b)$ is $\A$ irreducible. So suppose that
$m>0$, and set $B'=\frac{1}{\sqrt{m}}B$ so that
\ben B'^*B'=B'B'^*=Id.\een
Since $B'e=(1-e)B'$ we see that $B'e$ provides a partial isometry implementing a
Murray-von Neumann equivalence
\ben (B'e)^*(B'e)=e\qquad (B'e)(B'e)^*=(1-e).\een
Now we have established that
\ben \mu=2[(\A,e\HH,e\D e)]\in K^*(\A\otimes\A^{op}),\een
from which we may conclude that $\mu$ does not satisfy Poincar\'{e} Duality, since
for any generator $q\in K_*(\A)$, $q\cap [(\A,e\HH,e\D e)]$ 
can not be in the image of $\cap\mu$. Hence we have a
contradiction, and the representation of $\Omega^*_\D(\A_b)$ must be irreducible. }
\end{theorem}
\noindent
Before we can prove the Morita equivalence of $\A$ and $\Omega^*_\D(\A)$, we need 
a few results to help us identify 
our various algebras precisely. Recall that an ideal $I$ in a
$C^*$-algebra $A$ is called essential if the intersection of $I$
with every other ideal in $A$ is nonzero. In turn
this is equivalent to $aI=Ia=\{0\}\Rightarrow a=0$.

\begin{lemma} Suppose that $I$ is an ideal in the
  unital $C^*$-algebra $A$. Then
  $M_n(I)$ acts irreducibly on $A^n$ if and only if $I$ is an
  essential ideal of $A$.

\medskip
\proof{Suppose that $M_n(I)$ acts on $A^n$, and that
  $I$ is not essential. Then there exists $a\in A$ such that
  $aI=\{0\}=Ia$, and so $a$ commutes with every $b\in I$,
  contradicting irreducibility. 
\smallskip\newline
Conversely, suppose that $I$ is essential in $A$. Then $M_n(I)$
is essential in $M_n(A)$, and as $M_n(A)$ acts irreducibly on
$A^n$, we may suppose without loss of generality that $n=1$. Now
suppose that $I$ does not act irreducibly on $A$. Then there
exists a projection $p\in A$ such that $A=pA\oplus (1-p)A$ as an
$I$-module, and $pb=bp$ for all $b\in I$. As $I\cdot I\subseteq
I$, $I$ must be contained in one of the two halves. Suppose that
it is $pA$. Then $pb=bp=b$, and so $(1-p)b=0=b(1-p)$,
contradicting $I$ essential. Similarly, if $I\subseteq (1-p)A$ we
have $(1-p)b=b$ for all $b\in I$ and so $pb=bp=0$, and we again
obtain a contradiction.}
\end{lemma}

\begin{corollary} If $A$ is unital, $E=pA^n$ is finite
  projective, and $I$ is an ideal in $A$, the
representation
of $pM_n(I)p$ is irreducible if and only if $I$ is an 
essential
ideal in $A$.

\medskip
\proof{ This follows from the above by replacing $a\in M_n(A)$ by $pap$.}
\end{corollary} 
\noindent
We can now state our main result.

\begin{theorem} If $(\A,\HH,\D,c,J)$ is a spin manifold, with $\HH_\infty=p\A^n$, 
then $\Omega^*_\D(\A_b)\cong
pM_n(\A_b)p$. Furthermore, 
$\Omega^*_\D(\A)=pM_n(\A)p=End^0_\A(\HH_\infty)$, so $\A$ and 
$\Omega^*_\D(\A)$ are
strongly Morita equivalent. 

\medskip
\proof{As $\Omega^*_\D(\A_b)$ is unital and acts irreducibly on
  $\HH_\infty$,
\be \Omega^*_\D(\A_b)\cong pM_n(B)p\ee
where $B\subseteq \A_b$ is a unital subalgebra. This follows
because

1) $\A_b$ is unital and so is its own multiplier algebra, and 

2) $\Omega^*_\D(\A_b)$ acts irreducibly, and so must comprise a
full matrix algebra.

However, $\Omega^*_\D(\A_b)$ is also an $\A_b$ bimodule, so 
$\A_b\Omega^*_\D(\A_b)\subseteq\Omega^*_\D(\A_b)$. Thus $B$ must
be an ideal, and as it is unital, $B=\A_b$. 
\medskip\newline
The same argument applies to $\Omega^*_\D(\A)$ except that now
$B$ can be a proper ideal of $\A_b$. By the above results, it must be an
essential ideal, and since $\Omega^*_\D(\A)$ is also an
$\A$-bimodule, we have $\Omega^*_\D(\A)\cong pM_n(\A)p$.
\medskip\newline
As two algebras $\A$, $\B$ are (pre) strongly Morita equivalent if and only if 
$\B\cong End_\A^0(E)$ for some (pre) $C^*$ $\A$-module $E$, we have shown 
that $\A$ and $\Omega^*_\D(\A)$ are strongly Morita equivalent, with $\HH_\infty$
providing an equivalence bimodule.}
\end{theorem}

\begin{corollary} With $(\A,\HH,\D)$ as above, $\Omega^*_\D(\A_b)$ is finite projective 
(left or right) $\A_b$-module,
while $\Omega^*_\D(\A)$ is an $\A_b$-finite projective module.
\end{corollary}

 From what we 
have shown about the module $p\A^n$ we may also conclude that $p\A_b^n$ provides a
strong Morita equivalence between $\A_b$ and $\Omega^*_\D(\A_b)$. Thus in
noncommutative (spin) geometry the algebra $\Omega^*(\A)$ plays a r\^{o}le strongly
analogous to that of the Clifford algebra in the commutative case. In particular, any
triple $(\A,\HH,\D)$ satisfying the axioms listed above provides a noncommutative
analogue of a spin$^c$ structure, \cite{RP}.

A further problem is determining when a noncommutative Riemannian geometry, as
discussed in \cite{SL}, has a spin$^c$ structure. This amounts to identifying a
noncommutative analogue of the (first few) Stieffel-Whitney classes. Other problems
then arise, such as the compatibility of the two structures and  the relation to 
spin/Real structures. This will be the subject of a future paper.

\section{Acknowledgements} I would like to thank Steven Lord for drawing this problem to
my attention, and for several interesting conversations on the subject. I would also like
to thank Alan Carey for his support and interest. 

Adam Rennie
\newline
University of Adelaide
\newline
North Terrace, Adelaide, 
\newline
South Australia, 5005 
\newline
email: arennie@maths.adelaide.edu.au

\end{document}